%% file: ms.tex
\documentclass[runningheads]{llncs}
\usepackage[numbers,sort&compress]{natbib}

\usepackage{etoolbox}

\usepackage{todonotes}

\usepackage{graphicx}

\usepackage{amsmath}
\usepackage{amsfonts}
\usepackage{booktabs}
\usepackage{multicol}
\usepackage{multirow}

\usepackage{orcidlink}

\newcommand\norm[1]{\left\lVert#1\right\rVert}

\newcommand\orcid[1]{\textsuperscript{\orcidlink{#1}}}

\usepackage{xcolor}

\begin{document}
\title{MURR: \underline{M}odel \underline{U}pdating with \underline{R}egularized \underline{R}eplay for Searching a Document Stream}
\titlerunning{Model Updating with Regularized Replay}

\author{Eugene Yang\orcid{0000-0002-0051-1535}\inst{1} \and 
Nicola Tonellotto\orcid{0000-0002-7427-1001}\inst{2} \and
Dawn Lawrie\orcid{0000-0001-7347-7086}\inst{1} \and 
Sean MacAvaney\orcid{0000-0002-8914-2659}\inst{3} \and \\
James Mayfield\orcid{0000-0003-3866-3013}\inst{1} \and
Douglas W. Oard\orcid{0000-0002-1696-0407}\inst{4,1} \and
Scott Miller\orcid{0009-0003-3345-6346}\inst{5}
}

\authorrunning{E. Yang et al.}
\institute{
HLTCOE. Johns Hopkins University, Baltimore, Maryland, USA \\
\email{\{eugene.yang,lawrie,mayfield\}@jhu.edu}
\and 
University of Pisa, Pisa, Italy \email{nicola.tonellotto@unipi.it} 
\and
University of Glasgow, Glasgow, UK \email{sean.macavaney@glasgow.ac.uk}
\and
University of Maryland, College Park, Maryland, USA \email{oard@umd.edu} 
\and
Information Sciences Institute, University of Southern California, California, USA \email{smiller@isi.edu}
}

\maketitle              %
\begin{abstract}
The Internet produces a continuous stream of new documents and user-generated queries.
These naturally change over time based on events in the world and the evolution of language. 
Neural retrieval models that were trained once on a fixed set of query-document pairs will quickly
start misrepresenting newly-created content and queries, leading to less effective retrieval. 
Traditional statistical sparse retrieval can update collection statistics
to reflect these changes in the use of language in documents and queries.
In contrast, continued fine-tuning of the language model underlying neural retrieval approaches such as DPR and ColBERT
creates incompatibility with previously-encoded documents.
Re-encoding and re-indexing all previously-processed documents can be costly.
In this work, we explore updating a neural dual encoder retrieval model without reprocessing past documents in the stream. 
We propose MURR, a model updating strategy with regularized replay,
to ensure the model can still faithfully search existing documents without reprocessing,
while continuing to update the model for the latest topics. 
In our simulated streaming environments, we show that fine-tuning models using MURR
leads to more effective and more consistent retrieval results than other strategies
as the stream of documents and queries progresses.

\end{abstract}

\input{1-intro}

\input{2-background}

\input{3-method}

\input{5-experiment}
\input{6-results}

\input{7-conclusion}

\subsubsection{\ackname}
This work was partially supported by
the Spoke ``FutureHPC \& BigData'' of the ICSC – Centro Nazionale di Ricerca in High-Performance Computing, Big Data and Quantum Computing funded by the Italian Government, the FoReLab and CrossLab projects (Departments of Excellence), the NEREO PRIN project (Research Grant no. 2022AEFHAZ) funded by the Italian Ministry of Education and Research (MUR).

\bibliographystyle{splncs04nat}
\bibliography{bibio}

\end{document}

%% file: 1-intro.tex
\section{Introduction}

As a large amount of content is created every day,
search engines that aim to serve queries targeting old, new, or even emerging topics need to incorporate new documents continuously.  
Such a \textit{stream of documents} can address a fluid set of topics,
and in news, for example, we would expect that documents created at different times would discuss different issues. 
Public-facing search engines serving ad hoc queries, such as Baidu, Bing or Google,
need to account not just for topic changes in their document streams,
but also topical changes in their query streams over time~\cite{jatowt2024news, willis2016makes}.
For example, a query of ``Harris'' in 2024 likely targets a different ``Harris'' than was being searched for in 2014.
Fine-tuning a query-document relevance model---such as a BERT cross-encoder~\cite{monobert} or a ColBERT dual encoder~\cite{colbert}---only once would not work well in such cases;
some approach to periodic, incremental, or continuous updating is needed to reflect these changes.

These kinds of distributional shifts are less of a problem for earlier lexical retrieval systems, such as BM25, 
which can easily update their statistics as term distributions change;
they rely on directly-observable term statistics such as document frequency
or (for unigram language models) collection frequency.
As a result, lexical retrieval systems such as Lucene can easily incorporate emerging topics in a stream of documents. 
However, providing similar capabilities for dense retrieval systems,
such as those that use FAISS~\cite{faiss}, 
PLAID's cluster-based dot product approximation~\cite{santhanam2022plaid}, or other nearest-neighbor indexes,
is much more challenging. 
Popular approximate nearest neighbor (ANN) search algorithms 
(e.g., HNSW~\cite{malkov2018efficient} and Product Quantization~\cite{jegou2010product}))
support fast vector searches and
allow incremental ingestion of vectors to an index.
However, the parameters of the models that provide the vectors require training to reflect distributional changes.
Given that these models produce vectors in an arbitrary latent space,
an updated model is not guaranteed to be compatible with vectors from another model
(though they can be moderately compatible by chance~\cite{li2021encoder}).
Therefore, existing approaches either need to use the same model over the lifespan of a dense index (limiting effectiveness)
or re-index the entire collection when a new model is trained (a costly endeavor).
We address these limitations in this work.

There has been some work on this problem of nonstationary query and collection statistics under dense retrieval
in which the document and query encoders are fixed,
and adaptation to a new domain or genre uses few-shot training~\cite{beir}.
In this paper, we broaden that perspective to explore model updating strategies that adapt to emerging topics
through larger-scale fine-tuning on newly-collected training triples.
Since models trained on different training examples provide no guarantees
that existing documents will be represented using similar vectors,
we also study the effect of distributional shifts on our ability to retrieve older documents without reindexing.

In practical applications, we expect that search engines might use implicit feedback to obtain the training signals
that are needed to support model updating in the presence of changing distributional statistics~\cite{radlinski2005evaluating}.
In this work, 
we %
simulate document and query streams from an existing multi-domain retrieval collection to experiment with various streaming scenarios.

We introduce MURR, a \underline{m}odel \underline{u}pdating strategy with \underline{r}egularized training data \underline{r}eplay on their document representations. 
By assuming distributional shifts between known domains,
we can compare the effect of training data replay against that of representation regularization. 
We show that dense retrieval models trained with MURR are more effective than models without replay or representation regularization, and that they provide more consistent retrieval effectiveness to the same set of queries
 as distributional statistics change. 

Our contributions are threefold: 
    (1) an effective model updating strategy that does not require reprocessing existing documents,
    (2) a set of simulated query and document streams, and
    (3) experimental results demonstrating MURR's effectiveness and stability.

%% file: 2-background.tex
\section{Background and Related Work}

Neural retrieval models are generally based on a pretrained language model, such as BERT~\cite{bert} or RoBERTa~\cite{roberta}, and further fine-tuned with a retrieval objective, such as contrastive~\cite{infonce, chen2020simple} or distillation loss~\cite{colbertv2, formal2022distillation, lin2023prod}, on a set of queries and documents, such as MS MARCO~\cite{msmarco}. 
For dual encoders, because of their architecture, queries and documents can be processed separately and only require a lightweight interaction, such as vector dot product~\cite{dpr, splade} or token level MaxSim~\cite{colbert}; documents can thus be preprocessed and indexed for supporting fast searches~\cite{santhanam2022plaid, faiss}. 
How well the language model is tuned for the specific query and document language distributions is crucial for designing an effective dense retrieval model~\cite{lassance2023experimental}. 

Since proper training data is difficult to collect and the fine-tuning process is also computationally expensive, prior work in ad hoc retrieval typically aims for a one-size-fits-all general-purpose model and evaluates its robustness on various retrieval domains and tasks~\cite{wang2022text, ni2021large, formal2022distillation}.  This perspective drove the creation of the BEIR benchmark~\cite{beir}. 
Work on domain adaptation either aims for better zero-shot ability~\cite{thakur2022domain, chen2022out, zhuang2024setwise}, which is the same as developing a general-purpose model, or it exploits additional unsupervised or semi-supervised fine-tuning~\cite{wang2021gpl, hashemi2023dense, gangi2021synthetic}. 
However, with a stream of queries and documents arriving, even if one can afford to continuously fine-tune the retrieval model, the risk of catastrophic forgetting increases~\cite{lovon2021studying, goodfellow2013empirical}, which destroys the purpose of adaptation. 

As mentioned in the previous section, statistical sparse retrieval models such as BM25 are capable of incrementally updating the index and related statistics with some careful lexical feature selection to account for the language drift over time~\cite{lukes2018sentiment, florio2020time}. 
Early work around TREC filtering task, which dates back to TREC 4~\cite{harman1996overview}, also requires monitoring the lexical differences over time for serving standing queries~\cite{allan1996incremental}. 
In contrast, there is little work on neural dual encoders in such a streaming scenario. 
Recent work by \citet{plaid-shirttt} investigated incrementally indexing a large corpus using PLAID-X~\cite{tdistill} with a fixed model, but only evaluated on a fixed set of queries at the end of the stream. 
How the model performs during and across the stream remains unknown. 
Furthermore, updating the underlying model to adapt to emerging topics and language distributions is definitely beneficial for retrieval effectiveness, but it can be challenging. 
The recent LongEval CLEF Lab~\cite{galuscakova2023longeval} studies a similar problem but with queries on a set of persistent (standing) topics rather than a query stream that shifts topically over time. While closer to our setting, that setup limits us to studying cases in which the query topic distribution is stable.

%% file: 3-method.tex
\section{MURR: Model Updating with Regularized Replay}

In this section, we describe the training losses used for
\underline{m}odel \underline{u}pdating with \underline{r}egularized training data \underline{r}eplay -- MURR. 
The objective is to avoid reprocessing previously-indexed documents
while adapting the retrieval model to newly-arrived documents and queries
with backward compatibility that allows queries to target both new and old documents. 

Without loss of generality, we break the continuous stream of documents and queries into discrete periods,
named \textit{sessions}. 
Modeling and operational decisions are made between sessions but not within them.

In each session, we continue to fine-tune the trained retrieval model from the previous session (MURR-CF). 
The trained model is then used to encode and index the documents that arrive during the session,
resulting in multiple indexes encoded by \textit{different} models throughout the document stream.
As an alternative, one can fine-tune from a language model for each session (MURR-LM), which inherits 
no knowledge from the previous session (discussed more in Section~\ref{sec:exp:model}).
We use the model trained at the current session to encode incoming queries and search each existing vector index. 
Regularized replay, which is discussed in this section,
ensures that the models trained later in time are still compatible with existing indexes created by previous models.

\subsection{Contrastive Loss}

The typical training paradigm for dense retrieval models is contrastive learning~\cite{infonce},
which is often expressed using the SimCLR loss~\cite{chen2020simple}. 
For each training query $q_s\in\mathcal{Q}_s$ in the current session $s\in\mathbb{N}^0$,
we randomly sample one relevant and one non-relevant document,
denoted $d^+_{s}$ and $d^-_{s}$ respectively,
from the training collection to form the training triple $\left(q_s, d^+_s, d^-_s\right)\in \mathcal{D}_s$. 
With the triples, we formulate the contrastive loss for training the model $\mathcal{M}_s$ as
\begin{align}
    \mathcal{L}_{C}(\mathcal{D}_s) =  \frac{1}{|\mathcal{D}_s|} \sum_{\left(q, d^+,d^-\right)\in\mathcal{D}_s} -\log 
    \frac{
        \exp\left(f(\mathbf{q}_{(s)},  \mathbf{d}_{(s)}^+)\right)
    }{
        \exp\left(f(\mathbf{q}_{(s)},  \mathbf{d}_{(s)}^+)\right) + \exp\left(f(\mathbf{q}_{(s)},  \mathbf{d}_{(s)}^-)\right)
    }
\end{align}
where $\mathbf{q}_{(s)}$ and $\mathbf{d}_{(s)}$ denote the vector representations
encoded by the model $\mathcal{M}_s$ of the query and document, respectively,
and the function $f$ denotes the similarity function between the query and document vectors. 
In this work, we use the transformer model's embedding of the \texttt{[CLS]} token
as the vector representation of the text sequence;
we discuss this in detail in the next section. 

\subsection{Representation Regularized Replay}
At the end of training in each session $i$,
we sample a subset of training triples $\tilde{\mathcal{D}}_i\subset\mathcal{D}_i$,
along with their vector representations encoded by the trained model $\mathcal{M}_i$,
for memory replay in the future sessions through simple random sampling. 
At the current session $s$,
we construct an aggregated replay training set $\mathcal{D}^R_s = \cup_{i\in[0, s-1]} \tilde{\mathcal{D}_i}$.

For replay triples, we construct an additional representation regularization loss
to ensure that the current model $\mathcal{M}_s$ is still capable of matching queries
with existing document vectors produced by past models
by encouraging the current model to recreate the same document representation as was produced by the model that originally indexed the document.  
Specifically, the regularization loss is defined as
\begin{align}
    \mathcal{L}_{R}(\mathcal{D}^R_s) = \frac{1}{2|\mathcal{D}^R_s|} 
        \sum_{(q, d^+, d^-)\in\mathcal{D}^R_s}
        \norm{ \mathbf{d}_{(s)}^+ - \mathbf{d}_{(i)}^+}_2 + \norm{ \mathbf{d}_{(s)}^- - \mathbf{d}_{(i)}^-}_2
\end{align}
where $\norm{\cdot}_2$ denotes the $L^2$ norm of a vector. 

Therefore, the overall training loss of for model $\mathcal{M}_s$
is the combination of the contrastive loss on both the training triples from the current session and the replay history along with the regularization; this can be expressed as
\begin{align}
    \mathcal{L}_{s} = \mathcal{L}_C( \mathcal{D}_s \cup \mathcal{D}^R_s ) + \alpha\mathcal{L}_{R}(\mathcal{D}^R_s) 
\end{align}
where $\alpha$ is a hyperparameter for regularization strength. 

\subsection{Continue Indexing and Distributed Search}

At each session, we use model $\mathcal{M}_s$ to encode and index the documents coming in at session $s$. 
Each document is encoded into a vector and indexed into the approximate nearest neighbor search index $\mathcal{I}_s$. 

At search time, queries are encoded by the current model $\mathcal{M}_s$. 
We search all existing indexes, i.e., $\mathcal{I}_0, \mathcal{I}_1, ... \mathcal{I}_{s-1}$,
and merge the resulting ranked lists by document scores. 
With this approach, we do not need to retain or serve any previously trained model,
triage the queries to different models,
or reprocess any documents. 
In the remainder of the paper, we discuss the compatability of models trained using regularized replay
with document vectors encoded by different models.

%% file: 5-experiment.tex
\section{Experiments}

In this section, we describe our experiment setup for evaluating the proposed MURR strategy. 

\input{4-simulation}

\subsection{Dense Retrieval Model}\label{sec:exp:model}

We use Tevatron~\cite{tevatron} to train DPR~\cite{dpr} models with a feed-forward layer attached to the \texttt{[CLS]} token
to project the final hidden layer to a vector of 768 dimensions for the experiments. 
At each session, we \underline{c}ontinue to \underline{f}ine-tune the model trained in the previous session
using our proposed regularized replay (MURR-CF).
At Session 0, which is the very first session,
we initialize the DPR model with a retrieval-prepared language model
trained with CoCondenser~\cite{cocondenser}.\footnote{\url{https://huggingface.co/Luyu/co-condenser-marco}}
As a variant, we also evaluate a strategy that always initializes the model from the CoCondenser model
instead of the one from the previous session (MURR-LM). 
We use all training triples from the emerging domain (training set) at each session to train the model,
which is a different set from the set of queries and documents that the model would be indexing and searching on (the test set).
The statistics of the dataset are summarized in Table~\ref{tab:lotte_stats}.
At the end of each session, we retain 200 replay triples. 
We fine-tune the model for 20,000 gradient steps with an effective batch size of 256 using 8 NVIDIA V100 GPUs.
We optimize the model with an Adam optimizer,
a learning rate of $3\times10^{-6}$,
and a regularization strength of 0.01.  

Each document arriving during the current session is encoded with the latest DPR model, which is produced differently when using MURR-CF and MURR-LM, into a 768-dimension vector.
The vectors are indexed using FAISS~\cite{faiss} with product quantization~\cite{jegou2010product}
using a 4-bit fast scan that relies on SIMD instructions for distance computations. 
At search time, the queries are also encoded by the latest model
but distributed to all existing indexes to support searching documents that arrived in previous sessions. 
Ranked lists from indexes are combined based on the similarity scores of each document. 
Since no document is indexed more than once,
we do not need fusion approaches
(e.g., reciprocal rank fusion or min-max normalization). 

\subsection{Baselines}

Under the streaming setup, there are three baseline strategies that we compare with the proposed MURR:
\begin{itemize}
    \item \textbf{Same Model}.
    In most prior studies and applications, the underlying DPR model is not updated,
    and the previously-trained model continues to encode newly-arrived queries and documents. 
    \item \textbf{LM without Replay}.
    At the other extreme, the DPR model is always up-to-date by training only on the latest emerging domain.
    This strategy biases the model toward the latest topics but will be less effective in serving past ones. 
    \item \textbf{CF without Replay}.
    Similar to the previous one but continuously fine-tune the DPR model,
    which retains the knowledge learned from the previous sessions
    but may quickly be washed out by fine-tuning focused on the latest emerging domain. 
\end{itemize}

\subsection{Evaluation}\label{sec:exp:evaluation}

We evaluate the retrieval effectiveness by Success@5, which is 1 if at least one relevant document is at or above rank five and 0 otherwise,
as used in the work that introduced LoTTe~\cite{colbertv2}. 
To assess the significance of the effectiveness differences between strategies, we report the macro-average of Success@5 and use a paired $t$-test on Success@5 values on each query set across all sessions with a 95\% confidence level. 
Such significance tests essentially ask the question, ``Is the effectiveness of the resulting retrieval models when using strategy A different from using strategy B over the entire stream?''

However, since the focus of this work is not only on the absolute effectiveness of the model
but also the quality of retrieval from session to session,
quantitatively, we also report the relative difference between two consecutive sessions of the average Success@5 on a set of queries. 
This relative metric provides indicators of how well the retrieval model serves \textit{each set of search queries}
(realized as query sets in this work) throughout the stream with additional documents and updated retrieval models. 
In our experiments, the query set evaluated in each session is a union of the query sets from prior sessions and the new query set. %
This allows us to observe the effectiveness of queries targeting previously indexed documents and
documents indexed by the latest model.
Specifically, the relative Success@5 gain is calculated by
\begin{align}
    \frac{2}{S(S-1)}\sum_{s=0}^S \sum_{i=0}^{s-1} \frac{\text{Success@5}(\mathcal{Q}_i \mid \cup_{j=0}^{s} \mathcal{I}_j)}{\text{Success@5}(\mathcal{Q}_i \mid \cup_{j=0}^{s-1} \mathcal{I}_j)}
\end{align}
where $S$ is the number of sessions (five in our experiments),
and Success@5$(\mathcal{Q}_i \mid \cup_{j=0}^{s} \mathcal{I}_j)$ indicates the Success@5 value
evaluated on the test queries in Session $i$ with the accumulated indexes up to Session $s$. 
To assess the difference between the relative Success@5 gain,
we also report the standard deviation of the $\frac{S(S-1)}{2}$ values that we average over,
which we argue is more informative than merely reporting significance levels. 

This metric does \textit{not} consider how effective the model is on a specific query set;
instead, we measure how much the model \textit{retains} effectiveness throughout the stream. 
Theoretically, a model can perform extremely poorly on a set of queries and continue to do so,
so under this gain measure, the value would be higher than a model that performs very effectively on the set
but only slightly degrades as the stream progresses. 
Therefore, this gain measure is orthogonal to the effectiveness measure and evaluates the consistency of the strategy in serving a set of queries over time. 

%% file: 4-simulation.tex
\subsection{Dataset and Stream Simulation}\label{sec:simulation}

\input{_lotte_stats} 

We simulate the query and document streams using the forum subset of the LoTTe dataset~\cite{colbertv2}.
The LoTTE forum subset contains documents and queries from five domains,
with each domain split into training (named \textit{dev} in the original paper) and test sets.
The number of documents in each domain varies, while the number of queries per domain is roughly equal.
The statistics of the dataset are summarized in Table~\ref{tab:lotte_stats}. 

\subsubsection{Sampling Distribution. }

\begin{figure}[t]
    \centering
    \includegraphics[width=\linewidth]{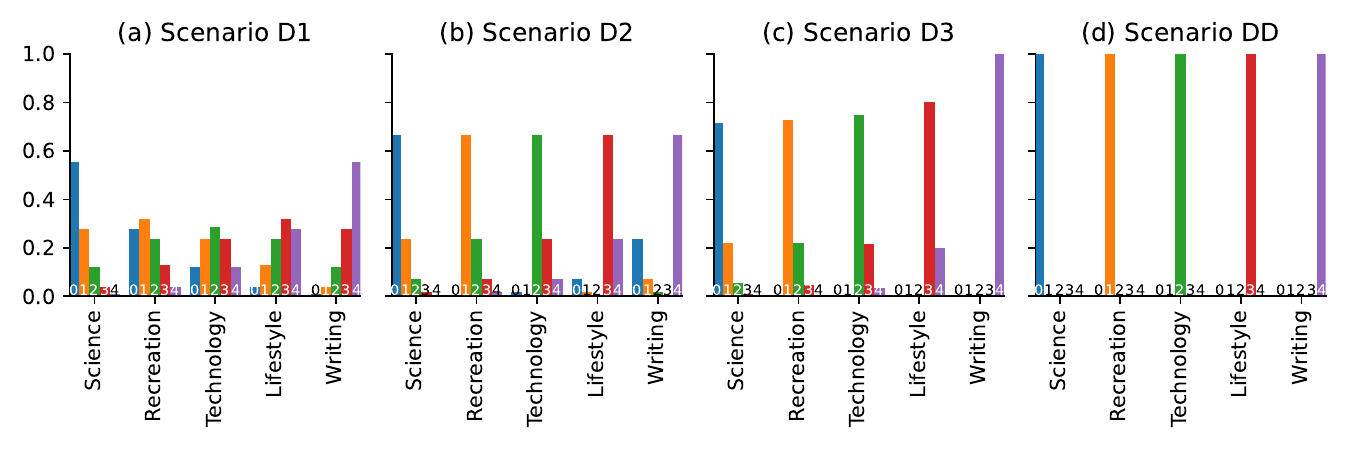}
    \caption{Sampling distribution of each domain in each session and simulated stream. The color of the bars represents the session, along with the session ID marked at the bottom of each bar.}
    \label{fig:simulation-distribution}
\end{figure}

For our experiments, we simulate four streams of data.
Each simulated data stream is split into five sessions,
representing five model or index updating decision points;
this is similar to prior work in document streaming~\cite{plaid-shirttt}. 
Each domain follows a beta-binomial sampling distribution over the sessions for both queries and documents.
The sampling distributions are summarized in Figure~\ref{fig:simulation-distribution}.
We start by sampling the queries.
Each session contains an emerging domain to simulate the topical shift of search queries over time. 
The ordering of the emerging domain is the same across all four simulated streams:
science, followed by recreation, technology, lifestyle, and writing.

We assume that search topics follow aspects of the world that are reflected in the document stream,
so in this work
we simulate
the same topical shift.
This is implemented by having %
documents follow the same sampling distribution as queries across sessions. 
However, for evaluation, documents relevant to a specific query
should appear in the current or earlier session in which the query was sampled.
This creates a sampling constraint on that document, so document sampling is not purely random. 
Therefore, to ensure this quality, a document relevant to a query follows a truncated sampling distribution
at the time of the session in which its query was sampled. 
Recall that while the number of queries in each domain is similar,
the number of documents in each domain is not.
Therefore, when sampling documents,
the number of documents will not mirror the distribution, while the number of queries does.
Thus, the resulting numbers of queries and documents in each session will naturally differ.  

\subsubsection{Scenario Simulated. }

Each simulated stream represents a different streaming scenario.
\begin{itemize}
    \item \textbf{D1} contains a constant flow of technology queries and documents,
    with a symmetric and smooth distribution peaking at session 2,
    while other domains shift in different sessions with a skewed distribution. 
    This simulates a mixture of different topical shifts in a combined stream. 
    
    \item \textbf{D2}. Each domain in D2 follows a five-session decay with a different starting point
    and rotates to the front of the sessions,
    illustrated in Figure~\ref{fig:simulation-distribution}(b). 
    While the documents follow the same sampling distribution,
    since the number of documents varies across domains,
    the sampled number of documents in each session is not proportional to the number of sampled queries. 
    This scenario simulates a seasonal shift in topics, 
    with one topic dominating a session but continuing to appear in the rest of the stream.
    Session 0 is, therefore, clipped from an infinite seasonal stream. 

    \item \textbf{D3} simulates a finite stream with each domain following a decaying distribution.
    Since the stream ends at Session 4, the length of the decay also decreases as the sessions
    progress, leaving the last domain (writing) to appear only in Session 4. 

    \item \textbf{DD} (where the second D indicates ``Default''), simulates discrete topic shifts,
    with each session containing only one domain.
    Despite being artificial, this stream is diagnostic
    for investigating the effectiveness and compatibility of model updating strategies.  
\end{itemize}

The script for creating the simulated scenarios and the specific simulated query and document streams used in this work
are released on Huggingface.\footnote{\url{https://huggingface.co/datasets/hltcoe/lotte-streams-for-murr}}

%% file: _lotte_stats.tex
\begin{table}[t]
\setlength\tabcolsep{0.55em}

\caption{Dataset Statistics of LoTTe Forum Datasets.}\label{tab:lotte_stats}
\centering

\begin{tabular}{cc|rrrrr}
\toprule
Split &     &  Science &  Recreation &  Technology &  Lifestyle &  Writing \\
\midrule
\multirow{2}{*}{Train}
& Documents &   343,642 &     263,025 &   1,276,222 &    268,893 &  277,072 \\
& Queries   &     2,013 &       2,002 &       2,003 &      2,076 &    2,003 \\
\midrule
\multirow{2}{*}{Test}
& Documents & 1,694,164 &     166,975 &     638,509 &    119,461 &  199,994 \\
& Queries   &     2,017 &       2,002 &       2,004 &      2,002 &    2,000 \\
\bottomrule
\end{tabular}

\end{table}

%% file: 6-results.tex
\section{Results and Analysis}

\begin{figure}[t]
    \centering
    \includegraphics[width=\linewidth]{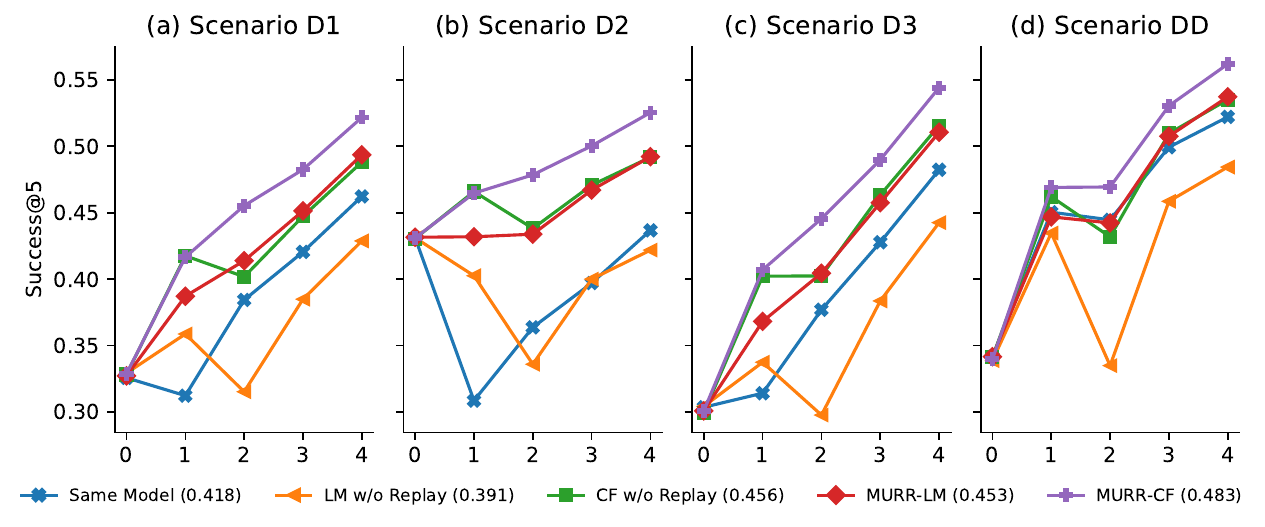}
    \caption{Success@5 of each model updating strategy where the x-axis is the running session ID. Values in parenthesis in the legend are macro-average of Success@5 across all query sets and streams. Differences between each strategy are all statistically significant using the paired t-test described in Section~\ref{sec:exp:evaluation} except for the pair of \textit{CF w/o Replay} and \textit{MURR-LM} after multiple test corrections.}
    \label{fig:main_results}
\end{figure}

Compared to other strategies, MURR-CF,
which continues to fine-tune the dense retrieval model from the previous session with regularized replay,
provides consistent improvements across sessions in all four simulated streams. 
Summarized in Figure~\ref{fig:main_results},
MURR-CF (purple with pluses) is statistically significantly more effective than other strategies
across all sessions except the very first one
(where all strategies result in the same model for that session).
We show that our proposed simple regularization on memory replay
provides a significant boost compared to only continued fine-tuning of models from the previous session
(CF w/o Replay, in green with squares).

Among the five strategies, training only on the latest training data
(LM w/o Replay, in orange with triangles)
is consistently and statistically significantly the least effective;
it is even worse than not updating the model
(Same Model, in blue with crosses),
which is the obvious baseline for a streaming scenario.
Only training on the latest data loses compatibility with previous sessions,
making old indexes incompatible with the current query representation. 
However, to our surprise, the representations are not totally incompatible.
Even in Scenario DD, where domain shifts are dramatic,
LM w/o Replay still achieves reasonable Success@5 values
(within 0.2 of the best in most cases)
despite being the worst among others. 

Continued fine-tuning without training data replay
(CF w/o Replay)
performs similarly to MURR-LM (red with diamonds),
which always initializes the DPR model with the CoCondenser model during fine-tuning with replay data.\footnote{We conduct additional equivalence test between \textit{CF w/o Replay} and \textit{MURR-LM} using two one-sided t-tests (TOST) with a 5\% relative difference as the threshold and conclude that the two strategies are statistically equivalent with 95\% confidence.}
The two share similar trajectories throughout the sessions in all streams except for the second session.
This indicates a small amount of training data replay with regularization on the document representations
is sufficient to replicate training from the previous session.
Combining both, which is MURR-CF, leads to more effective models than either alone,
indicating that, although similar in net effect, they improve the model in different ways. %

\subsection{Effectiveness on Individual Query Sets}

\begin{figure}[t]
    \centering
    \includegraphics[width=\linewidth]{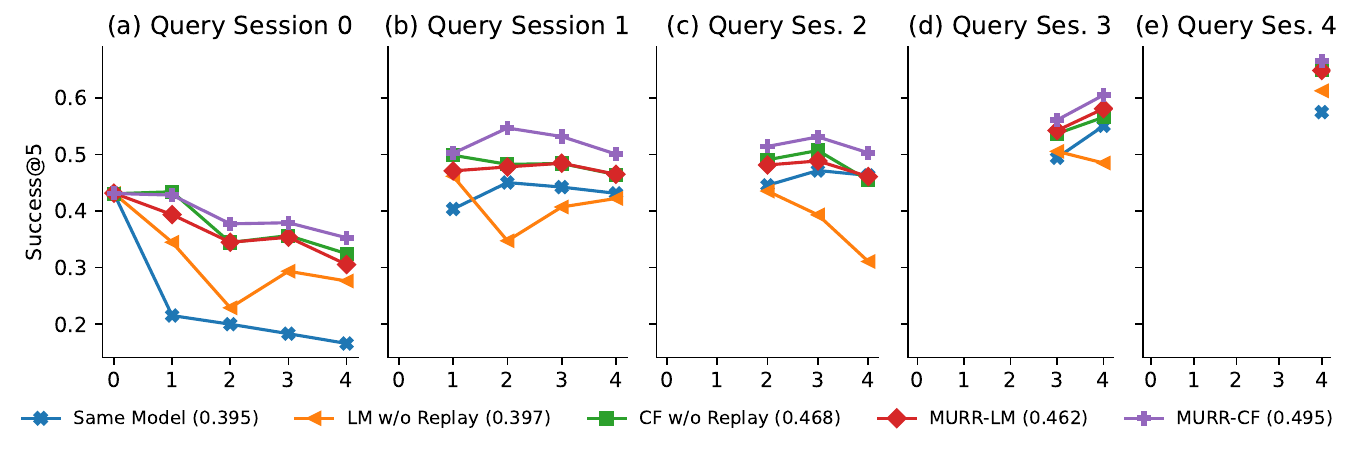}
    \caption{Query set breakdown of five model updating strategies in Scenario D2. Each subgraph is the effectiveness over sessions of the set of queries introduced in the session indicated in the title. Values in parenthesis are macro-average Success@5 on query sets in only D2. Values at the x-axis is the running session ID. }
    \label{fig:d2_breakdown}
\end{figure}

For a specific set of queries, MURR-CF is also more effective than other strategies
in terms of both absolute effectiveness and stability from session to session. 
As shown in Figure~\ref{fig:d2_breakdown}, MURR-CF is consistently more effective in all five query sets in Scenario D2.
Due to space constraints, we only present graphs on Scenario D2,
but we observe similar relationships between strategies in other simulated scenarios despite the differences in stream characteristics. In fact, all four simulated scenarios demonstrate similar strategy orders, while some exhibit larger variations in absolute effectiveness than others. 
Interestingly, in almost all cases, the Success@5 of the second session after the query set is introduced 
is higher than the session when it first appeared,
but that measure gradually decreases as the stream progresses.
This is likely an artifact of the stream simulation
since the sampling probability of a domain is similar between the peak and the very next session.  

\input{_main_relative_gain}

To characterize the consistency of retrieval quality from session to session,
we measure the relative Success@5 gain,
as described in Section~\ref{sec:exp:evaluation}.
Summarized in Table~\ref{tab:relative_gain}, on average, all strategies maintain roughly the same effectiveness
after their introduction to the stream,
with MURR-CF being the only strategy providing positive gains.
This indicates that, on average, effectiveness is improving with additional off-domain training.
While the other strategies offer negative gains, they all maintain roughly the same level of effectiveness. 
However, in terms of stability, LM w/o Replay,
which has no modeling connections from session to session,
is the least stable strategy with the highest standard deviation (overall 0.30).
With more explicit connections, such as continued fine-tuning and regularized replay,
MURR-CF becomes more stable (overall standard deviation 0.10).

Again, we observe similar effects between CF w/o Replay and MURR-LM
in both means and standard deviations over the four simulated scenarios,
with MURR-LM being slightly more stable than CF w/o Replay.
This indicates that the model memory likely has a limit
and will eventually move on from knowledge acquired during the early sessions that are no longer relevant to later sessions.
However, since our experimental setup only simulates a short stream,
it cannot clearly demonstrate this effect.

\subsection{Ablation Studies}

\begin{figure}[t]
    \centering
    \includegraphics[width=\linewidth]{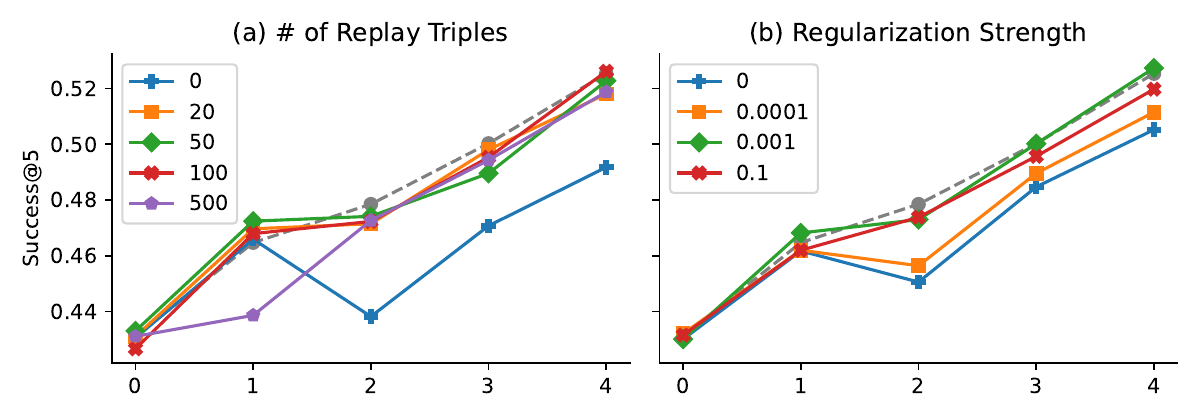}
    \caption{Ablation study on the number of replay triples and the regularization strength under MURR-CF in Scenario D2.
    The x-axis is the session ID. 
    The dashed gray line is the setup (200 triples, $\alpha=0.01$) used in the main experiments.
    Using 0 replay triples is essentially \textit{CF w/o Replay} in the main results. }\label{fig:ablation}
\end{figure}

We study the robustness of the two hyperparameters we introduced in MURR-CF:
number of replay triples and regularization strength. 
Summarized in Figure~\ref{fig:ablation},
we observe that MURR-CF is robust to both the number of replay triples and the regularization strength,
although the existence of both is still critical.
Using smaller or stronger regularization strength than 0.01 results in slight but noticeable degradation.
Using 0 or 0.0001 as the regularization strength, essentially only replaying triples but no representation regularization,
leads to less effective results.
Particularly for Session 2, both using no replay and no regularization
led to a degradation in effectiveness from the previous session;
using some replay triples and a moderate regularization strength provided a stable trajectory. 
Surprisingly, while more replay triples provide stability to the process,
a small number of replay triples are sufficient to calibrate the models trained in the following sessions
to be compatible with document representations from the previous sessions. 
However, the 0 replay triples case is significantly worse than any runs that use some replay.
We observe a very similar gap in Scenario D1 and much smaller gaps in Scenarios D3 and DD, suggesting that such representation regularization depends on the stream, which we leave for future investigation.

%% file: _main_relative_gain.tex
\begin{table}[t]
\caption{Relative Success@5 Gain. Reported values are macro-averaged over the query sets sampled in each session (for example, the queries sampled for Session 0 have four gains over the subsequent four sessions, but the queries sampled for session 1 have three), as discussed in Section~\ref{sec:exp:evaluation}.  The standard deviation is in parentheses.}\label{tab:relative_gain}
    
\centering
\resizebox{\linewidth}{!}{
\setlength\tabcolsep{0.45em}
\begin{tabular}{c|r|rr|rr}
\toprule
Scenario &     Same Model &  LM w/o Replay &  CF w/o Replay &        MURR-LM &        MURR-CF \\
\midrule
D1     &   0.046 (0.17) &   0.049 (0.32) &   0.041 (0.20) &   0.043 (0.17) &   0.055 (0.18) \\
D2     &  -0.052 (0.17) &  -0.070 (0.19) &  -0.034 (0.08) &  -0.030 (0.07) &  -0.013 (0.07) \\
D3     &  -0.032 (0.15) &  -0.020 (0.42) &  -0.016 (0.12) &  -0.023 (0.09) &   0.005 (0.07) \\
DD     &  -0.008 (0.01) &  -0.065 (0.27) &  -0.046 (0.07) &  -0.040 (0.06) &  -0.020 (0.02) \\
\midrule
Overall&  -0.012 (0.14) &  -0.027 (0.30) &  -0.014 (0.13) &  -0.012 (0.11) &   0.007 (0.10) \\
\bottomrule
\end{tabular}
}
\end{table}

%% file: 7-conclusion.tex
\section{Conclusion and Future Work}

This work proposes a strategy for updating a neural retrieval model
that adapts to changes in document topicality over time
without requiring documents already seen to be reindexed.
We simulated four scenarios that contain streams of queries and documents
to evaluate our proposed MURR model updating strategy. 
We showed that our proposed method provides more effective and more stable retrieval results throughout the streams
without reprocessing the existing documents. 
These results demonstrate the feasibility of deploying a dense retrieval model
and continuing to modify it on-the-fly for emerging topics. 

While simulation experiments provide critical evidence for developing such model-updating strategies,
experimenting on real document streams with real language distribution shifts can further empirically verify our findings.
We plan to evaluate on LongEval to verify our findings in this work. 
Additionally, collecting documents from a stream, such as Common Crawl,
and developing queries at different snapshots (or assigning existing queries to appropriate time points) is also in our plan to evaluate evolving information needs over time. 